\documentclass[11pt]{article}
\usepackage{epsf} 
\usepackage{amsmath}
\usepackage{amssymb}
\usepackage{epsfig}
\usepackage{latexsym}
\usepackage{amsfonts}
\usepackage{graphicx}%
\usepackage{varioref}
\usepackage{ifthen}
\begin{document}
\parindent 0mm 
\setlength{\parskip}{\baselineskip} 
\thispagestyle{empty}
\pagenumbering{arabic} 
\setcounter{page}{0}
\mbox{ }
\rightline{UCT-TP-290/2012}
\newline
\rightline{MZ-TH/12-32}
\newline
\rightline{September 2012}
\newline
\vspace{0.1cm}

\begin{center}
\long\def\symbolfootnote[#1]#2{\begingroup%
\def\thefootnote{\fnsymbol{footnote}}\footnote[#1]{#2}\endgroup}

\LARGE \textbf{Corrections to the ${\bf SU(3)\times SU(3)}$ Gell-Mann-Oakes-Renner relation and chiral couplings $L^r_8$ and $H^r_2$}
\LARGE \symbolfootnote[1]{{\LARGE {\footnotesize Supported in part by  MEC and FEDER (EC) under grant FPA2011-23596 and GV under grant PROMETEO2010-056 and BEST12, by NRF (South Africa), and by the Alexander von Humboldt Foundation (Germany).}}}
\end{center}
\begin{large}
\begin{center}
{\bf J. Bordes}$^{(a)}$,
{\bf  C. A. Dominguez} $^{(b)}$,
{\bf  P. Moodley} $^{(b)}$, 
{\bf J. Pe\~{n}arrocha}$^{(a)}$, 
and {\bf K. Schilcher} $^{(b),(c)}$ \\
\end{center}
\end{large}
\begin{center}
$^{(a)}$Departamento de F\'{\i}sica Te\'{o}rica,
Universitat de Valencia, and Instituto de F\'{\i}sica Corpuscular, Centro
Mixto Universitat de Valencia-CSIC\\
\vspace{.4cm}
$^{(b)}$Centre for Theoretical \& Mathematical Physics, University
of Cape Town, Rondebosch 7700, South Africa\\
\vspace{.4cm}
$^{(c)}$Institut f\"{u}r
Physik, Johannes Gutenberg-Universit\"{a}t, Staudingerweg 7, D-55099
Mainz, Germany
\end{center}

\begin{center}
\textbf{Abstract}
\end{center}
\noindent
Next to leading order corrections to the $SU(3) \times SU(3)$ Gell-Mann-Oakes-Renner relation (GMOR)  are obtained using weighted QCD Finite Energy Sum Rules (FESR) involving the pseudoscalar current correlator. Two types of integration kernels in the FESR are used to suppress the contribution of the kaon radial excitations to the hadronic spectral function, one with local and the other with global constraints. The result for the pseudoscalar current correlator at zero momentum is $\psi_5(0) = (2.8 \pm 0.3) \times 10^{-3} \;{\mbox{GeV}}^4$, leading to the chiral corrections to GMOR: $\delta_K = (55 \pm 5)\%$. The resulting uncertainties are mostly due to variations in the upper limit of integration in the FESR, within the stability regions, and to a much lesser extent due to the uncertainties in the strong coupling and the strange quark mass. Higher order quark mass corrections, vacuum condensates, and the hadronic resonance sector play a negligible role in this determination. These results confirm an independent determination from chiral perturbation theory giving also  very large corrections, i.e. roughly an order of magnitude larger than the corresponding corrections in chiral $SU(2) \times SU(2)$. Combining these results with our previous determination of the corrections to GMOR in chiral $SU(2) \times SU(2)$, $\delta_\pi$, we are able to determine two low energy constants of chiral perturbation theory, i.e. $L^r_8 = (1.0 \,\pm \, 0.3)\, \times 10^{-3}$, and $H^r_2 = - (4.7 \,\pm \, 0.6) \, \times 10^{-3}$, both at the scale of the $\rho$-meson mass.

\newpage

\section{Introduction}
The light quark condensate, an order parameter of chiral-symmetry breaking, is known to play an important role in QCD as well as in chiral perturbation theory (CHPT) \cite{PAGELS}. In particular, the ratio of the strange to the up/down quark condensate, $R_{sq}=\langle\bar{s}s\rangle/\langle\bar{q}q\rangle$, is a key QCD quantity measuring flavour $SU(3)$ symmetry breaking in the vacuum \cite{CAD0}-\cite{DNS1}, and entering in many QCD sum rule applications (for a review see \cite{REVIEW}). This ratio is also related to two CHPT low energy constants, which in turn determine the next-to-leading order corrections to the Gell-Mann-Oakes-Renner (GMOR) relation \cite{PAGELS},\cite{GMOR}
\begin{equation}
\psi _{5}(0)|_{i}^{j}\,\equiv \,-(m_{i}+m_{j})\left\langle 0 \right\vert 
\overline{q_i}q_i+\overline{q_j}q_j\left\vert 0 \right\rangle\, = \,2\,f_P^2\,M_P^2\,(1\,-\,\delta_P)\;,
\end{equation} 
where $i,j$ stand for (light) quark flavours, $f_P$ and $M_P$ are respectively the decay constant and the mass of the pion or kaon, with  
$f_K/f_\pi = 1.197 \pm 0.006$, $f_\pi = 92.21 \pm 0.14\;\mbox{MeV}$ \cite{PDG}-\cite{ROSNER}, and $\delta_P$ is the chiral correction. The relation between $R_{sq}$ and the CHPT low energy constants $L_8^r$ and $H_2^r$ is \cite{BERN}-\cite{SCHERER}
\begin{equation}
R_{sq}\equiv \frac{<\bar{s} s>}{<\bar{q} q>} = 1 + 3 \mu_\pi - 2 \mu_K - \mu_\eta + \frac{8}{f_\pi^2} (M_K^2 - M_\pi^2) (2 L^r_8 + H^r_2) \; ,
\end{equation}
where $<\bar{q} q>$ is the average of the up- and down quark condensates, $L^r_8$ is a (physical) low-energy constant in the  chiral Lagrangian to next-to-leading order, and 
\begin{equation}
\mu_P = \frac{M_P^2}{32 \pi^2 f_\pi^2} \ln \frac{M_P^2}{\nu_\chi^2} \;,
\end{equation}
with $\nu_\chi$ the chiral renormalization scale. The constant $L^r_8$ at a scale of the rho-meson mass has been estimated in chiral perturbation theory to next-to-leading order  with the result \cite{JAMIN1}
\begin{equation}
L^r_8(\nu_\chi = M_\rho) = (0.88 \pm 0.24) \times 10^{-3} \;,
\end{equation}
while a determination at order $\cal{O}$$(p^6)$ gives \cite{P6} $L^r_8(\nu_\chi = M_\rho) = (0.62 \pm 0.20) \times 10^{-3}$. The unphysical low energy constant $H^r_2$ has been estimated in \cite{JAMIN1} as
\begin{equation}
H^r_2 (\nu_\chi = M_\rho) = - (3.4 \pm 1.5) \times 10^{-3} \;.
\end{equation}
These CHPT low energy constants are related to the next-to-leading order chiral corrections to GMOR as follows
\begin{equation}
\delta_\pi = 4 \frac{M_\pi^2}{f_\pi^2} ( 2 L^r_8 - H^r_2) \;\;\;\; and\;\;\;\; \delta_K = \frac{M_K^2}{M_\pi^2} \,\delta_\pi \;.
\end{equation}
Using the results Eqs.(4)-(5) in Eq.(6) the following estimate was obtained in \cite{JAMIN1}
\begin{equation}
\delta_\pi = (4.7 \pm 1.7)\% \;\;\;\;\; and \;\;\;\;\; \delta_K = (61 \pm 22)\% \;.
\end{equation}
It is also possible to determine these corrections to GMOR entirely from QCD sum rules involving the correlator of the axial-vector current divergences  at zero momentum. The feasibility of such an approach has been established long ago \footnote{In this respect there is a technical issue regarding logarithmic quark mass singularities which will be addressed in the next section.} in a series of papers \cite{BROAD0}-\cite{CAD1}.
Recently, such an independent determination of $\delta_\pi$  from weighted QCD Finite Energy Sum Rules (FESR), and in good agreement with the value above, has been obtained in \cite{CAD2} with the result
\begin{equation}
\delta_\pi = (6.0 \pm 1.0)\%  \;.
\end{equation}
Regarding $\delta_K$, the CHPT result in Eq.(6) already indicates that the chiral $SU(3) \times SU(3)$ corrections to GMOR are expected to be huge; in fact, much larger than typical chiral corrections such as the $10-15 \%$ deviation from the $SU(3) \times SU(3)$ Goldberger-Treiman relation \cite{GTR}. Hence, it is important to confirm such a huge correction by determining it independently of the CHPT result in Eq.(6). In this paper we perform such a determination using modern weighted QCD FESR \cite{DNS1}, \cite{CAD2}, \cite{DNS2}-\cite{DNS3} designed to reduce considerably the systematic uncertainties arising from a lack of direct experimental information on the pseudoscalar hadronic spectral function. In addition, by combining the result for $\delta_K$ with our previous determination  of $\delta_\pi$, Eq.(8), we are able to determine $L^r_8$ and $H^r_2$ independently, and entirely from QCD sum rules.

\section{Finite energy QCD sum rules}
The pseudoscalar current correlator $\psi_5(q^2)|_i^j$, for quark flavours $i=u$ and $j=s$, is defined as
\begin{equation}
\psi _{5}(q^{2})|_u^s\,=\,i\int \,d^4 x\,e^{iqx}<0|\,T(j_{5}(x)\,j_{5}(0))\,|0 >\;,
\end{equation}
where $<0|$ is the physical vacuum and the pseudoscalar current density $j_{5}(x)$ is
\begin{equation}
j_{5}(x)\,=\,(m_{s}+m_{u})\;:\overline{s}(x)\,i\,\gamma _{5}\,u(x):\ ,
\end{equation}
and $m_{u,s}$ are the light quark masses. At zero momentum, and using a Ward identity one obtains the GMOR relation \cite{PAGELS}, \cite{BROAD0}-\cite{BROAD1}
\begin{equation}
\psi _{5}(0)|_u^s\, \equiv \,-\left[m_{s}(\mu^2)+m_{u}(\mu^2)\right]\left\langle 0 \right\vert 
\overline{s}s+\overline{u}u\left\vert 0 \right\rangle(\mu^2)\, = \,2\,f_K^2\,M_K^2\,(1\,-\,\delta_K)\;,
\end{equation}
where $\mu$ is a renormalization scale. Using Wick's theorem in the time ordered product in Eq.(9) would  lead to normal-ordered operators in the low energy theorem, Eq.(11), and $\psi_5(0)$ would be a renormalization group invariant quantity. In principle, however, this is spoiled by the presence of mass-singular quartic terms in perturbative QCD(PQCD), as well as tadpole contributions, which would cancel out if the quark condensate becomes minimally subtracted instead of normal-ordered \cite{BROAD1}-\cite{CAD1} (for a recent reassessment see \cite{DNS1}). In practice, though, one can safely neglect quartic mass corrections in the case of the up- and down-quark correlator, and consider normal-ordered quark condensates so that there is no residual $\mu$-dependence in $\psi_5(0)|_u^d$. In the case of $\psi_5(0)|_u^s$, though,  it would appear a priori that quartic strange-quark mass corrections could be sizable. However, a posteriori, and given the  resulting uncertainties in the value of $\delta_K$ it turns out that one can also disregard this issue which then becomes purely academic.\\
The  FESR are obtained  by invoking Cauchy's theorem in the complex squared-energy plane. The pseudoscalar current correlator at zero momentum is then given by
\begin{equation}
\psi_5(0) \Delta_5(0) =  \frac{1}{\pi }\int_{s_{th}}^{s_0} \frac{\Delta_5(s)}{s}\,Im\,\psi _{5}(s)\; ds
\, +
\frac{1}{2\pi i}\oint_{C(|s_0|) } \,\frac{\Delta_5(s)}{s} \;\psi _{5}(s)
\,ds\,,
\end{equation}
where in the sequel we drop the quark-flavour indexes, $s_{th}\, =\,  M_K^2$ is the hadronic threshold, the integration path $C(|s_0|)$ is a circle of  radius $|s_0|$, and $\Delta_5(s)$ is an arbitrary analytic integration kernel to be described shortly. The radius $|s_0|$ determines the onset of the hadronic continuum beyond the resonance region, which is assumed to be given by PQCD. Hence, the first integral in Eq.(12) involves the hadronic spectral function, and the second integral the QCD correlator. The latter is known up to five-loop order in PQCD
\cite{CHET0}-\cite{CHET}. Mass terms up to order $m_s^6$, and leading order vacuum condensates with Wilson coefficients to two-loop order are also known \cite{CAD1}. A posteriori, these contributions turn out to be negligible on account of the overall uncertainty in $\psi_5(0)$ (roughly at the 10\% level), mostly due to variations inside the stability region in $|s_0|$, and to a lesser extent due to the uncertainty in the value of the strong coupling and the strange quark mass. 
The hadronic spectral function involves the kaon pole followed by its radial excitations, i.e.
\begin{equation}
\frac{1}{\pi} \, Im \;\psi _{5}(s)\,=\, 2\, f_K^2 \,M_K^4\, \delta(s - M_K^2)\; + \; \frac{1}{\pi} \, Im \;\psi _{5}(s)|_{RES} \;,
\end{equation}
where the first two resonances, $K_1 \equiv K(1460)$, and $K_2 \equiv K(1830)$, have masses and widths known from experiment, i.e. $M_{K_1}= 1460 \;{\mbox{MeV}}$, $\Gamma_{K_1}= 250 \;{\mbox{MeV}}$, $M_{K_2}= 1830 \;{\mbox{MeV}}$, $\Gamma_{K_2}= 250 \;{\mbox{MeV}}$ \cite{PDG}. 
This experimental information, though, is restricted to the values of the masses and the widths. Hence, a reconstruction of the resonance spectral function entirely from data is not feasible, leading to an unknown systematic uncertainty from model dependency. For instance, inelasticity and non-resonant background are realistically impossible to model. In an attempt to reduce  this model dependency, it was first proposed long ago \cite{CAD3}-\cite{CAD6} to normalize pseudoscalar spectral functions at threshold using  CHPT as a constraint. In the case of the kaonic channel, the CHPT constrained resonance spectral function, which includes the important resonant sub-channel $K^*(892) - \pi$, is given by \cite{CAD6} (there is a misprint in \cite{CAD6} which is corrected below)
\begin{equation}
\frac{1}{\pi} \; \mbox{Im} \; \psi_{5}(s)|_{K \pi \pi} =
\frac{M_{K}^{4}}{2f_{\pi}^{2}} \; \frac{3}{2^{8} \pi^{4}} \;
\frac{I(s)}{s (M_{K}^{2} - s)^2} \; \theta (s - M_{K}^{2}) \;,
\end{equation}
where
\begin{eqnarray*}
I(s) = \int_{M_{K}^{2}}^{s} \; \frac{du}{u} \; (u - M_{K}^{2}) \; (s - u) \;
\Biggl\{ (M_{K}^{2} - s) \left[ u - \frac{(s+M_{K}^{2})}{2} \right] \Biggr.
\end{eqnarray*}
\begin{equation}
\Biggl. - \frac{1}{8u} \; (u^{2} - M_{K}^{4}) \; (s - u) + \frac{3}{4} \;
(u - M_{K}^{2})^{2} |F_{K^{*}} (u)|^{2} \Biggr\} \; ,
\end{equation}
and
\begin{equation}
|F_{K^{*}} (u)|^{2} = \frac{ \left[ M_{K^{*}}^{2} - M_{K}^{2} \right]^{2} +
M_{K^{*}}^{2} \; \Gamma_{K^{*}}^{2}}
{ (M_{K^{*}}^{2} - u)^{2} + M_{K^{*}}^{2} \; \Gamma_{K^{*}}^{2}} \; . 
\end{equation}
The pion mass has been neglected above, in line with the approximation
$m_{u} = 0$ to be made in the QCD sector, and the resonant sub-channel $\rho(770) - K$ is numerically negligible. The complete hadronic spectral function is then
\begin{eqnarray*}
\frac{1}{\pi} \; \mbox{Im} \; \psi_{5}(s)|_{HAD} =
2 f_{K}^{2} M_{K}^{4} \;\delta (s - M_{K}^{2}) +
\frac{1}{\pi} \; \mbox{Im} \; \psi_{5}(s)|_{K \pi \pi} 
\frac{[BW_{1}(s) + \lambda \;BW_{2}(s)]}{(1 + \lambda)}
\end{eqnarray*}
\begin{equation}
+ \; \frac{1}{\pi} \; \mbox{Im} \; \psi_{5}(s)|_{PQCD} \;\theta (s - s_{0})\; ,
\end{equation}
where $\mbox{Im} \; \psi_{5}(s)|_{QCD}$ is
the perturbative QCD spectral function modeling the continuum which
starts at some threshold $s_{0}$, $BW_{1,2}(s)$ are Breit-Wigner forms
for the two kaon radial excitations, normalized to unity at threshold,
and $\lambda$ controls the relative importance of the second radial
excitation. The choice $\lambda \simeq 1$ results in a reasonable
(smaller) weight of the K(1830) relative to the K(1460). While further embellishments to this parameterization could still be made, they would not hide its strong model dependency and its related unknown systematic uncertainty. This problem has recently been solved by a judicious choice of the integration kernel, $\Delta_5(s)$, in the FESR \cite{DNS1}, \cite{CAD2}, \cite{DNS2}-\cite{DNS3}. In fact,  choosing a second degree polynomial constrained to vanish at the peaks of the two radial excitations makes this contribution to the FESR completely negligible in comparison with the uncertainties due to the value  of $\alpha_s$, the strange quark mass, and variation of the results in the stability region in $|s_0|$. In the kaon channel the expression for  $\Delta_5(s)$ is 
\begin{equation}\label{eq:Delta5}
\Delta_5(s) = 1 - a_0 \, s - a_1 \; s^2 \;
\end{equation}
with $\Delta_5(M_1^2)= \Delta_5(M_2^2)= 0$, which fixes $a_0= 0.768 \;{\mbox{GeV}^{-2}}$, and 
$a_1= - 0.140 \;{\mbox{GeV}^{-4}}$. Other choices of kernels with local constraints do not lead to any appreciable improvement in the quenching of the resonant contribution to the FESR. The optimal choice of integration kernel is most likely application dependent. For instance, pinched kernels in FESR for the vector and axial-vector correlators
\cite{PINCH1}-\cite{PINCH3} improve considerably the agreement between data and the first two Weinberg sum rules. This circumstance, often referred to as duality violation, is a contentious issue as some model analyses claim its presence \cite{DV1}-\cite{DV2}, and other model independent analyses claim the opposite \cite{DV3}. In the present case, a combination of Eq.\eqref{eq:Delta5} and pinched kernels does not work as well as Eq.\eqref{eq:Delta5} on its own \cite{DNS3}. However, it is possible to achieve some degree of pinching using other type of kernels as described next.\\

A different kernel with a global, rather than local constraint, has been used recently in connection with the corrections to GMOR in chiral SU(2), and it involves Legendre type polynomials \cite{CAD2} as explained next.
We consider the general polynomial
\begin{equation}
P_{n}(s)= \sum_{m=0}^n\ a_m \; s^{m}\; ,  
\end{equation}
with coefficients  fixed by (i) imposing the normalization condition at
threshold 
\begin{equation}
P_{n}\left( s=M_{K }^{2}\right) \,=\,1 \;,  \
\end{equation}
and (ii) by requiring that the polynomial $P_{n}(s)$  minimizes the
contribution of the continuum in the range $\left[ s^{(k+1)}_{\mathrm{th}},s_0
\right] $ in a least square sense, i.e.,
\begin{equation}
\int_{s^{(k+1)}_{\mathrm{th}}}^{s_0}\,\,s^{m} \, P_{n}(s)\,\,ds=0\,\,\;\;\mathrm{for}%
\,\,m=0,\ldots n-1 \;.  
\end{equation}%
The meaning of the threshold $ s^{(k+1)}_{\mathrm{th}}$ is as follows. The hadronic resonance spectral function can be split into the sum of a term involving the known kaon radial excitations, and a term involving  unknown resonances,  non-resonant background, etc., i.e.
\begin{equation}
Im \,\psi_5 ^{(k)}(s)\;=\; \sum_{i=0}^{k} Im \,\psi_{5}^{(i)}(s)+\theta (s-s^{(k+1)}_{
\mathrm{th}}) \;Im \,\psi _{5}^{(k+1)}(s)  \;,
\end{equation}
where the upper index $k$ stands for the number of resonances explicitly
included in the sum. The polynomials obtained in this way are closely related to the Legendre polynomials, so that Eq.(21) is exact on account of  orthogonality. Specifically, denoting by $\cal{P}$$_n(x)$ the Legendre polynomials, their relation to the $P_n(s)$ above is
\begin{equation}
P_{n}(s)\,=\,\frac{\mathcal{P}_{n}\left( x(s)\right) }{\mathcal{P}_{n}\left(
x(M_{K}^{2})\right) }  \;,
\end{equation}
where the variable $x(s)$ is
\begin{equation}
x(s)\,=\,\frac{2s\,-\,(s_0+s^{k+1}_{\mathrm{th}})}{s_0-s^{(k+1)}_{\mathrm{th}%
}}\;,
\end{equation}
defined in the range $x\,\in \,[-1,1]$ for  $s\,\in \,[s^{(k+1)}_{\mathrm{th}%
},s_0]$. The normalization of the $\cal{P}$$_n(x)$ is such that $\cal{P}$$_1(x) = x$, $\cal{P}$$_2(x) = (5 x^3 - 3x)/2$ , etc. The use of these polynomials leads to results for $\psi_5(0)$ in a region of large values of $s_0$ ($s_0 \, \simeq 20 -50 \;{\mbox{GeV}^2}$), where it is stable, and where PQCD is unquestionably valid. Since the Legendre polynomials exhibit zeros, this type of integration kernel provides  some amount of pinching. For instance, the sixth-degree polynomial vanishes at $\sqrt{s_0} \simeq 2.0\, {\mbox{GeV}}, 9.0 \, {\mbox{GeV}}$, and other four places at higher energies. 
\begin{figure}
[hb]
\begin{center}
\includegraphics[height=3.5in, width=5.0in]
{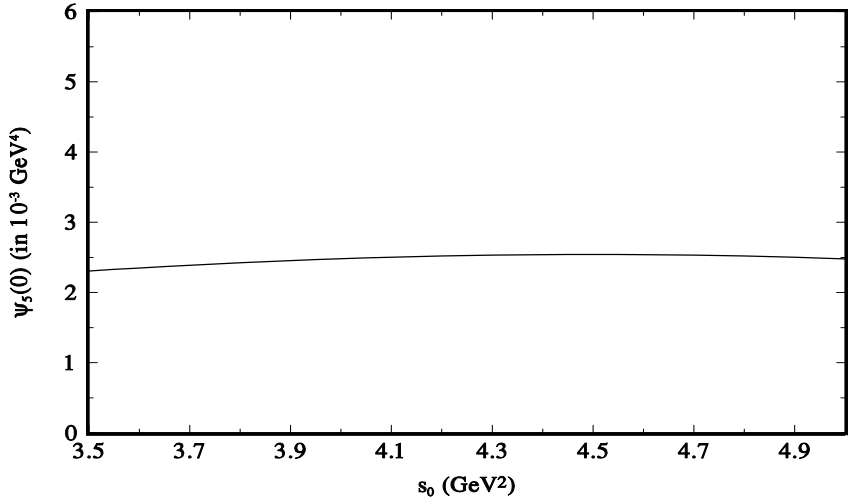}
\caption{Results for $\psi_5(0)$ in units of $10^{-3}\, \mbox{GeV}^4$ as a function of $s_0$, for    $\alpha_s(M_\tau^2) =0.326$ ($\Lambda\,=\, 350\; \mbox{MeV}$), and using the local integration kernel, Eq.(18).}
\end{center}
\end{figure}
\begin{figure}
[ht]
\begin{center}
\includegraphics[height=3.5in, width=5.0in]
{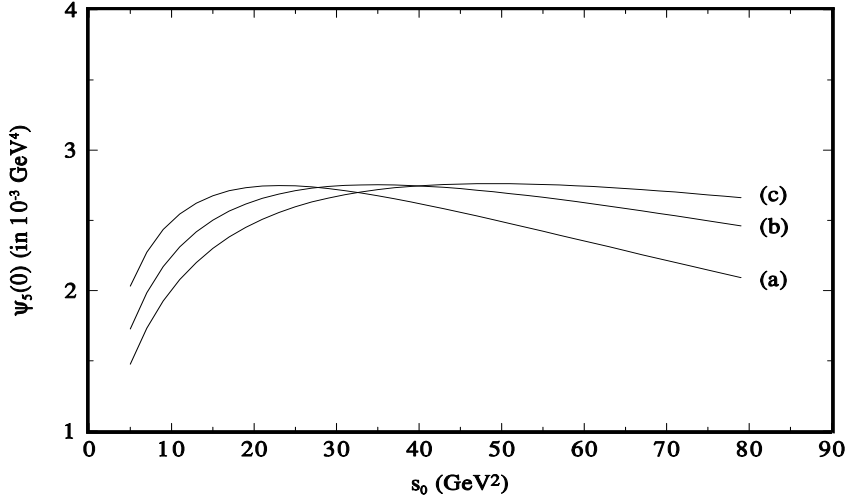}
\caption{Results for $\psi_5(0)$ in units of $10^{-3}\, \mbox{GeV}^4$ as a function of $s_0$, for    $\alpha_s(M_\tau^2) =0.326$ ($\Lambda\,=\, 350\; \mbox{MeV}$). Curves (a), (b), (c) correspond, respectively, to Legendre-type integration kernels $P_4$, $P_5$, and $P_6$ in Eq.(19).}
\end{center}
\end{figure}
\section{Results and conclusions}
We consider first the FESR involving the locally constrained integration kernel, Eq.(18), and in Fixed Order Perturbation Theory (FOPT). In this case the strong running coupling and the quark masses at a scale $|s|=|s_0|$ are considered fixed, thus only logarithmic terms contribute to the integrals. The renormalization group summation of leading logarithms is carried out after integration by setting the scale to $\mu^2 = - |s_0|$, and all integrals can be done analytically. For the strong coupling we use the wide range $\overline{\alpha_s}(M_\tau) = 0.31 - 0.35$, which covers many different results as reported recently in \cite{WKSHP} (including the PDG average \cite{PDG}). Using this range of values of $\alpha_s$ to determine the strange-quark mass in a similar framework as here \cite{DNS2} gives $\overline{m_s}|_{\overline{MS}}(2 \, {\mbox{GeV}}) = 100 - 104 \; {\mbox{MeV}}$, in the $\overline{MS}$ scheme (for recent reviews see \cite{QMASSREVIEW1}-\cite{QMASSREVIEW2}). Alternatively, one could use results from lattice QCD (for a compilation see \cite{PDG}) most of which are in good agreement within errors with \cite{DNS2}. Combining both methods we consider the wide range $\overline{m_s}|_{\overline{MS}}(2 \, {\mbox{GeV}}) = 94 - 104 \; {\mbox{MeV}}$.
In Fig. 1 we show a typical result for $\psi_5(0)$ as a function of $|s_0|$ in the wide stability region $|s_0|= 3 - 6 \;{\mbox{GeV}^2}$, corresponding to $\overline{\alpha_s}(M_\tau) = 0.326$ ($\Lambda_{QCD}= 350 \;{\mbox{MeV}}$), and $\overline{m_s}|_{\overline{MS}}(2 \, {\mbox{GeV}}) = 100\; {\mbox{MeV}}$. Taking into account the uncertainties in $\alpha_s$ and $m_s$ leads to 
\begin{equation}
\psi_5(0) = (2.8 \pm 0.3) \times 10^{-3} \;{\mbox{GeV}}^4 \;,
\end{equation}
which translates into the chiral corrections to GMOR 
\begin{equation}
\delta_K = (55 \pm 5) \%\;.
\end{equation}
Proceeding to the second choice of integration kernel, i.e. the Legendre type polynomials, typical results for  $\psi_5(0)$ are shown in Fig. 2 for the same input values of $\alpha_s$ and $\overline{m_s}$. Stability is achieved in a very wide region already for a sixth-order polynomial with the result: $\psi_5(0) = (2.50 -  3.00) \times 10^{-3} \;{\mbox{GeV}}^4 \;$, in excellent agreement with the  FOPT determination, Eq.(25). The chiral correction, Eq.(26), confirms the  CHPT determination  of \cite{JAMIN1}. \\

We discuss in closing a determination of the two CHPT constants $L^r_8$, and $H^r_2$ at the scale of the $\rho$-meson mass. It was first pointed out in \cite{CAD85}, and later reiterated in \cite{CAD0} and \cite{DNS1}, that in order to obtain the ratio $R_{sq} \equiv \langle \bar{s} s \rangle/\langle \bar{q} q \rangle$ with some reasonable accuracy one should avoid deriving it from the ratio of $\psi_5(0)^s_u/\psi_5(0)^d_u$, as this leads to a huge uncertainty. Instead, one should first determine the ratio of the scalar to the pseudoscalar correlator at zero momentum,  $ R_{VA} \equiv \psi(0)^s_u/\psi_5(0)^s_u$, and then use $R_{sq} \simeq (1 + R_{VA})/(1 - R_{VA})$. The most recent determination in a framework similar to the one used here, i.e. using weighted QCD FESR, gives \cite{DNS1}
\begin{equation}
R_{sq} \equiv \frac{\langle \bar{s} s \rangle}{\langle \bar{q} q \rangle} \, = 0.6 \, \pm \, 0.1\;.
\end{equation}
Using this value in Eq.(2) determines $2 L^r_8 + H_2$, and from the first relation in Eq.(6) one obtains  $2 L^r_8 - H_2$. Using the result for $\delta_\pi$ in Eq.(8), obtained in the same framework as here, gives
\begin{equation}
L^r_8 = (1.0 \, \pm \, 0.3) \, \times \, 10^{-3} \;
\end{equation}
\begin{equation}
H^r_2 = - (4.7 \, \pm \, 0.6) \, \times \, 10^{-3} \;,
\end{equation} 
in very good agreement with \cite{JAMIN1}. The value of $L^r_8$ above is also in agreement within errors with recent determinations from lattice QCD (for a review see \cite{FLAVIA}): $L^r_8 = (0.62 \pm 0.04) \times 10^{-3}$ \cite{PACS}, $L^r_8 = (0.85 \pm 0.04) \times 10^{-3}$ \cite{RBC}, and $L^r_8 = (0.88 \pm 0.08) \times 10^{-3}$ \cite{MILC}, as well as with recent analytical determinations:
$L^r_8 = (0.6 \pm 0.4) \times 10^{-3}$ \cite{Rosell}, and $L^r_8 = (0.64 \pm 0.16) \times 10^{-3}$ \cite{Bijnens}.  


\end{document}